# N- and p-type carrier injections into WSe$_2$ with van der Waals contacts of two-dimensional materials

Yohta Sata[1], Rai Moriya[1,*], Satoru Masubuchi[1], Kenji Watanabe[2], Takashi Taniguchi[2], and Tomoki Machida[1,3,*]

[1]*Institute of Industrial Science, The University of Tokyo, Meguro, Tokyo 153-8505, Japan*

[2]*National Institute for Materials Science, Tsukuba, Ibaraki 305-0044, Japan*

[3]*Institute for Nano Quantum Information Electronics, The University of Tokyo, Meguro, Tokyo 153-8505, Japan*

We demonstrated n-type and p-type carrier injections into a transition metal dichalcogenide (TMD) WSe$_2$ using van der Waals (vdW) contacts of two-dimensional (2D) materials: graphite for an n-type contact and NbSe$_2$ for a p-type contact. Instead of conventional methods such as the evaporation of metals on TMD, 2D metals were transferred onto WSe$_2$ in order to form van der Waals contacts. With these contacts, we demonstrated a small Schottky barrier height for both carrier polarities. Our finding reveals the potential of a high-performance vdW metal/semiconductor contact for use in electronics applications.

*E-mail: moriyar@iis.u-tokyo.ac.jp; tmachida@iis.u-tokyo.ac.jp

## 1. Introduction

Transition metal dichalcogenide (TMD) is extremely attractive for electronics and optoelectronics applications [1-4]. In the bulk crystal, these materials have layered crystal structures; thus, individual layers have strong covalent bonding within the plane and different layers are vertically held together with van der Waals (vdW) interlayer force [5]. Because of the relatively weak vdW interaction, these materials can be easily exfoliated down to monolayer without having dangling bonds on the exfoliated surface. These unique TMD properties are highlighted for a wide variety of applications such as high-performance transistors [6,7], sensors [8], flexible devices [9], and photodetectors [10,11]. However, the large contact resistance between metal and TMD seriously restricts the performance of these TMD-based devices [12]. In general, the material of the metal contact for TMD is selected in such a way that the bottom of the conduction band of TMD and the work function of the metal are aligned to form a Schottky-barrier-free n-type contact. In contrast, the contact can be of the p-type when the top of the valence band of TMD and the work function of the metal are aligned. Previous studies proved that the contact at metal/TMD interfaces was not solely determined by the position of the metal's work function with respect to the conduction or valence band of TMD [10]. It has been considered that there is an interface state at the evaporated metal/TMD interface. Because of the interface state, the Fermi level of the metals is pinned within the midgap of TMD, resulting in a large contact resistance at the evaporated metal/TMD interface. Instead of the evaporation of metals, the van der Waals contact between the two-dimensional (2D) metal and TMD has been drawing attention lately [13,14]. Examples of 2D metals are graphene, 1T'-WTe$_2$, and 2H-NbSe$_2$, whereas conventional metals that are deposited by thermal evaporation such as Au, Ti, and Pd are referred to as three-dimensional (3D) metals. Owing to the nonbonding nature of the vdW contact, it is expected that the interface state at the 2D metal/semiconducting TMD will be suppressed; thus, the surface pinning effect of the metal's Fermi level can be reduced. Moreover, since 2D metals have a wide variety of work functions, a significant reduction in contact resistance for both n- and p-type contacts for TMD is expected. A new method of forming contacts on TMD provides an opportunity to deepen our understanding of the properties of TMD. Here, we study the effect of a vdW contact between 2D metal/WSe$_2$ interfaces by fabricating field-effect transistor devices.

The work functions of 2D and conventional 3D metals are illustrated in Fig. 1(a). The positions of the conduction and valence bands of monolayer WSe$_2$ are shown together. The corresponding values are obtained from the recent density functional theory calculation results presented in Ref. 13. Various 2D metals with different work functions are also shown. The lowest and highest work functions of 2D metals are comparable to those of 3D metals. In our experiment, we selected graphite as a small-work-function material and NbSe$_2$ as a large-work-function material [15,16]. Regarding both materials, there were reports on their exfoliation down to monolayer and the fabrication of high-quality vdW heterostructures with other 2D materials [17-28]. Thus, they are suitable for forming n- and p-type contacts on WSe$_2$.

## 2. Device fabrication

The fabricated device structure and device micrograph are shown in Figs. 1(b)−1(e) for the graphite/WSe$_2$ [Figs. 1(b) and 1(d)] and NbSe$_2$/WSe$_2$ [Fig. 1(c) and 1(e)] devices.

Device fabrication was conducted by the one-by-one dry transfer of 2D materials in an atmospheric environment without any heating procedure [27,29]. First, hexagonal boron nitride (h-BN) with a thickness of 20−30 nm was mechanically exfoliated and deposited on a 300 nm $SiO_2$/doped-Si substrate. Second, few-layer $WSe_2$ was transferred to the flat surface of h-BN. Third, 3- to 6-nm-thick graphite or $NbSe_2$ layers were transferred. Finally, $WSe_2$ channels and part of the 2D metal contact were covered with another h-BN with a thickness of few nm to few tens of nm. The top and bottom h-BN layers served as the atomically flat substrate and encapsulation layer, respectively. These h-BN layers improved the uniformity of the $WSe_2$ layer and the adhesion between 2D metals and $WSe_2$ during transfer. The $WSe_2$ and $NbSe_2$ layers were exfoliated from bulk crystals (HQ Graphene Inc.). Graphite was exfoliated from Kish graphite. The electrical contacts for graphite and $NbSe_2$ were fabricated by the e-beam (EB) lithography and EB evaporation of 40 nm Au/40 nm Ti. The contact resistance of Au/Ti/graphite and Au/Ti/$NbSe_2$ junctions fabricated by this method was less than 1 kΩ and the junctions showed ohmic characteristics. The channel length $L$ and channel width $W$ of the devices were $W = 6.1$ μm and $L = 5$ μm for the graphite-contact device and $W = 3.5$ μm and $L = 4.5$ μm for the $NbSe_2$-contact device. The transport properties of the devices were measured in a variable-temperature cryostat from 240 to 300 K. Prior to the measurement, the devices were annealed for 6 h at 410 K within the cryostat. For transport measurements, the back-gate voltage $V_{BG}$ was applied between the device and the doped-Si substrate to control the carrier density of the $WSe_2$ layer. The source-drain bias voltage $V_{SD}$ was applied between the 2D metal contacts, and the source-drain current $I$ flowing through the device was measured.

## 3. Results and discussion

The transport properties at room temperature for both the graphite/WSe$_2$/graphite and NbSe$_2$/WSe$_2$/NbSe$_2$ devices are shown in Figs. 2(a)–2(f). The $V_{BG}$ dependence of the device current $I$ measured under constant $V_{SD}$ = 50 mV and 0.5 V, which are shown in Figs. 2(a) and (b), reveals distinct characteristics as follows. The graphite/WSe$_2$/graphite device exhibits a significantly higher current when the channel is electron-doped (positive $V_{BG}$ region) than when it is hole-doped (negative $V_{BG}$ region). In contrast, the NbSe$_2$/WSe$_2$/NbSe$_2$ device exhibits an opposite trend such that the current is higher when the WSe$_2$ channel is hole-doped than when it is electron-doped. From these comparisons, the graphite/WSe$_2$ vdW contact is preferred to be of the n-type, while the NbSe$_2$/WSe$_2$ vdW contact is preferred to be of the p-type. Since WSe$_2$ is exfoliated from the same batch of bulk crystal and the exfoliated flakes have nearly the same thickness, we think that the difference observed in Figs. 2(a) and 2(b) originates from the different carrier injection properties between graphite/WSe$_2$ and NbSe$_2$/WSe$_2$. We also note that the hysteresis between different $V_{BG}$ sweep directions is small for both devices because of h-BN encapsulation [30]; therefore, we think that the contamination of the WSe$_2$ surface caused by the adsorption of molecules could also be small for both devices. The current-voltage ($I$-$V_{SD}$) characteristics at $V_{BG}$ = +50 V (-50 V) are presented for graphite/WSe$_2$/graphite [Fig. 2(c)] and NbSe$_2$/WSe$_2$/NbSe$_2$ [Fig. 2(e)]. The small nonlinearlity observed in the $I$-$V_{SD}$ curves suggests the small Schottky barrier at the 2D metal/WSe$_2$ junctions. The field-effect mobility of the devices is determined by the relation $\mu_{FE} = [L/(WCV_{SD})](dI/dV_{BG})$, where $C = \varepsilon_0\varepsilon/d$ is the capacitance per unit area with the relative permittivity $\varepsilon$ of 3.9 for both SiO$_2$ and h-BN, $\varepsilon_0$ being the vacuum permittivity and $d$ the thickness of the dielectrics. By differentiating $I$-$V_{BG}$ curves, the mobilities for both devices are calculated at $V_{SD}$ = 0.6 V and presented in Figs. 2(d) and 2(f). The carrier mobilities extracted by this method monotonically increase with $V_{BG}$, suggesting that the two-terminal transport of each device is limited by the contact resistance at 2D metal/WSe$_2$ rather than by the channel resistance of WSe$_2$. Generally, the change in contact resistance with $V_{BG}$ induces a nonlinear $I$-$V_{BG}$ curve; thus, the mobilities are not constant with respect to $V_{BG}$. The significance of contact resistance could be inferred from the shift of the threshold voltage of the device shown in Figs. 2(a) and 2(b), which depends on $V_{SD}$. The nonlinearity in the $I$-$V_{SD}$ curve, which appeared in Figs. 2(c) and 2(e), also supports this speculation. Nevertheless, as a reference, we extracted the maximum electron mobility $\mu_e$ = 18 cm$^2$V$^{-1}$s$^{-1}$ for the graphite/WSe$_2$/graphite device and maximum hole mobility $\mu_h$ = 17 cm$^2$V$^{-1}$s$^{-1}$ for the NbSe$_2$/WSe$_2$/NbSe$_2$ device. With the assumption that the two-terminal transports of the devices are fully dominated by their contact resistance, we estimated the contact resistance of the devices using the following sequence. First, the drain current $I$ at $V_{SD}$ = +0.6 V was measured at $V_{BG}$ = +50 V for the graphite/WSe$_2$/graphite device and at $V_{BG}$ = -50 V for the NbSe$_2$/WSe$_2$/NbSe$_2$ device. $V_{SD}/I$ was nearly equal to the contact resistance 2$R_c$. Then, $R_c$ was multiplied by its contact width $W$. We obtained $R_cW$ = 780 kΩμm for the graphite/WSe$_2$ contact and $R_cW$ = 540 kΩμm for the NbSe$_2$/WSe$_2$ contact.

Next, to evaluate the Schottky barrier height between 2D metal and WSe$_2$, the temperature dependence of the $I$-$V_{BG}$ curve at $V_{SD}$ = 0.1 V was measured and the results are plotted in Fig. 3(a) for the graphite-contact device and in Fig. 3(d) for the NbSe$_2$-contact device. The lower the temperature, the lower the conductance in both devices;

this suggests that the thermionic emission across the metal/TMD interface is a dominant transport mechanism. These data were analyzed on the basis of the thermionic emission theory such that

$$I = A^*T^2 \exp\left(-\frac{e\varphi_B}{k_B T}\right)\left[\exp\left(-\frac{eV_B}{k_B T}\right)+1\right],\qquad(1)$$

where $A^*$ represents the effective Richardson constant, $\varphi_B$ the barrier height at the metal/WSe$_2$ interface, $e$ the elementary charge, $k_B$ the Boltzmann constant, and $T$ the temperature. By making the Arrhenius plot, the measured data is fitted with Eq. (1) and the derived barrier height $\varphi_B$ is plotted against $V_{BG}$ as shown in Figs. 3(b) and 3(e). $\varphi_B$ changes with $V_{BG}$ according to the gate modulation of the carrier density in the WSe$_2$ layer. There are particular $\varphi_B$ values indicated by arrows in Figs. 3(b) and 3(e); above these values, $\varphi_B$ changes linearly with respect to $V_{BG}$, reflecting thermionic transport, while below them, both thermionic and tunneling transport processes are reflected. It is generally known that these particular $\varphi_B$ values give the Schottky barrier height $\varphi_{SB}$ at a metal/TMD interface [31]. From these analysis results, we obtained $\varphi_{SB}$ values of 63 meV between the Fermi level of graphite and the conduction band edge of WSe$_2$ [(Fig. 3(c)] and 50 meV between the Fermi level of NbSe$_2$ and the valence band edge of WSe$_2$ [Fig. 3(f)]. Since the surface pinning effect at a 3D metal/WSe$_2$ interface pins the metal's Fermi level at 0.2–0.3 eV above the valence band [10], the demonstration of the n-type contact for WSe$_2$ with graphite suggests that the vdW contact is not restricted by the pinning effect. The obtained Schottky barrier height between graphite and WSe$_2$ is comparable to that obtained at a low resistance 3D metal/MoS$_2$ contact; it ranges from 20 to 100 meV [10]. Here, we compared our observation with the result of the 3D metal/MoS$_2$ contact since MoS$_2$ is more widely used for the n-type TMD transistor. Therefore, our finding suggests the potential advantage of using the graphite/WSe$_2$/graphite structure for n-type transistors. Notably, the p-type Schottky barrier height that we achieved in the NbSe$_2$/WSe$_2$ junction is significantly smaller than the typical values for a 3D metal/WSe$_2$ interface, which range from 0.27 to 0.45 eV [10]. Furthermore, such a junction is superior to recently reported NbSe$_2$/W$_x$Nb$_{1-x}$Se$_2$/WSe$_2$ vdW heterojunctions fabricated from CVD-grown TMD crystals [28,32]. We believe that our mechanical exfoliation of CVT-grown NbSe$_2$ and encapsulation with h-BN crystals provide a higher quality NbSe$_2$/WSe$_2$ vdW interface; thus, these methods enable us to achieve a smaller Schottky barrier height.

Our experimental results reveal the control between the n- and p-type contacts with a 2D metal/WSe$_2$ vdW heterostructure. This suggests that the vdW contact is not seriously restricted by the Fermi level pinning effect. However, the obtained Schottky barrier heights at graphite/WSe$_2$ and NbSe$_2$/WSe$_2$ are different from the theoretically calculated energy alignment presented in Fig. 1(a). From Fig. 1(a), it is expected that the graphite/WSe$_2$ junction will behave as an ambipolar contact with a large Schottky barrier height for both carriers, and that the NbSe$_2$/WSe$_2$ junction will behave as a p-type contact without a barrier. We note that the conduction and valence band positions of WSe$_2$ shown in Fig. 1(a) are calculated in the monolayer case, not the few-layer WSe$_2$ used in our experiment. However, the difference in band gap between the few-layer WSe$_2$ and the monolayer WSe$_2$ is ~0.2 eV [33,34], which is too small to explain the discrepancy. By using the band gap of the few-layer WSe$_2$ ($E_g$ ~ 1.40 eV) [33,34] and the obtained band offset at the graphite/WSe$_2$ and NbSe$_2$/WSe$_2$ interfaces in our experiment, the work function difference between graphite and NbSe$_2$ is calculated as ~1.29 eV. This difference in work function is close to the theoretical difference of 1.30 eV [Fig. 1(a)]. Therefore, we think that, by shifting the relative energy between WSe$_2$ and 2D metal, our experimental results and theoretical calculations exhibit a better agreement. It has been considered that interlayer coupling at a vdW junction alters the relative energy alignment between WSe$_2$ and metal [35,36]; this could explain our observation. More detailed comparisons between experimental and theoretical results by fabricating a series of 2D metal/TMD vdW junctions with different 2D metals are necessary in future experiments.

## 4. Conclusions

We demonstrated n- and p-type carrier injections into WSe$_2$ using vdW contacts between WSe$_2$ and 2D metals: graphite for an n-type contact and NbSe$_2$ for a p-type contact. The fabrication of n- and p-type transistors based on the same channel material could be suitable for circuit applications. This finding supports the concept of utilizing a vdW metal/semiconductor contact superior to those used in conventional methods.

## Acknowledgements

This work was supported by CREST, Japan Science and Technology Agency (JST) and JSPS KAKENHI Grant Numbers JP25107003, JP25107004, JP26248061, JP15H01010, and JP16H00982.

**Figure captions**

Fig. 1.

(Color online)(a) The theoretically calculated work functions of monolayer 2D and 3D metals are plotted together with the positions of the bottom of the conduction band and the top of the valence band of monolayer $WSe_2$. Their values were obtained from Ref. 13. (b, c) Structures of (b) graphite/$WSe_2$/graphite and (c) $NbSe_2$/$WSe_2$/$NbSe_2$ devices. (d, e) Optical microscopy images of (d) graphite/$WSe_2$/graphite and (e) $NbSe_2$/$WSe_2$/$NbSe_2$ devices.

Fig. 2.

(Color online) (a, b) $I-V_{BG}$ relationship of (a) graphite/$WSe_2$/graphite and (b) $NbSe_2$/$WSe_2$/$NbSe_2$ devices. The data measured in different $V_{BG}$ sweep directions are plotted. (c, e) $I-V_{SD}$ curves of (c) graphite/$WSe_2$/graphite ($V_{BG}$ = +50 V) and (e) $NbSe_2$/$WSe_2$/$NbSe_2$ ($V_{BG}$ = -50 V) devices. (d, f) $V_{BG}$ dependence of the carrier mobility in (d) graphite/$WSe_2$/graphite and (f) $NbSe_2$/$WSe_2$/$NbSe_2$ devices.

Fig. 3.

(Color online) (a, d) Temperature dependence of $I-V_{BG}$ curves of (a) graphite/$WSe_2$/graphite and (d) $NbSe_2$/$WSe_2$/$NbSe_2$ devices measured from 300 to 240 K ($V_{SD}$ = 0.1 V). (b, e) $V_{BG}$ dependence of the Schottky barrier height at (b) graphite/$WSe_2$ and (e) $NbSe_2$/$WSe_2$ junctions. (c, f) Illustration of obtained band offset at (c) graphite/$WSe_2$ and (f) $NbSe_2$/$WSe_2$ junctions.

Figure 1

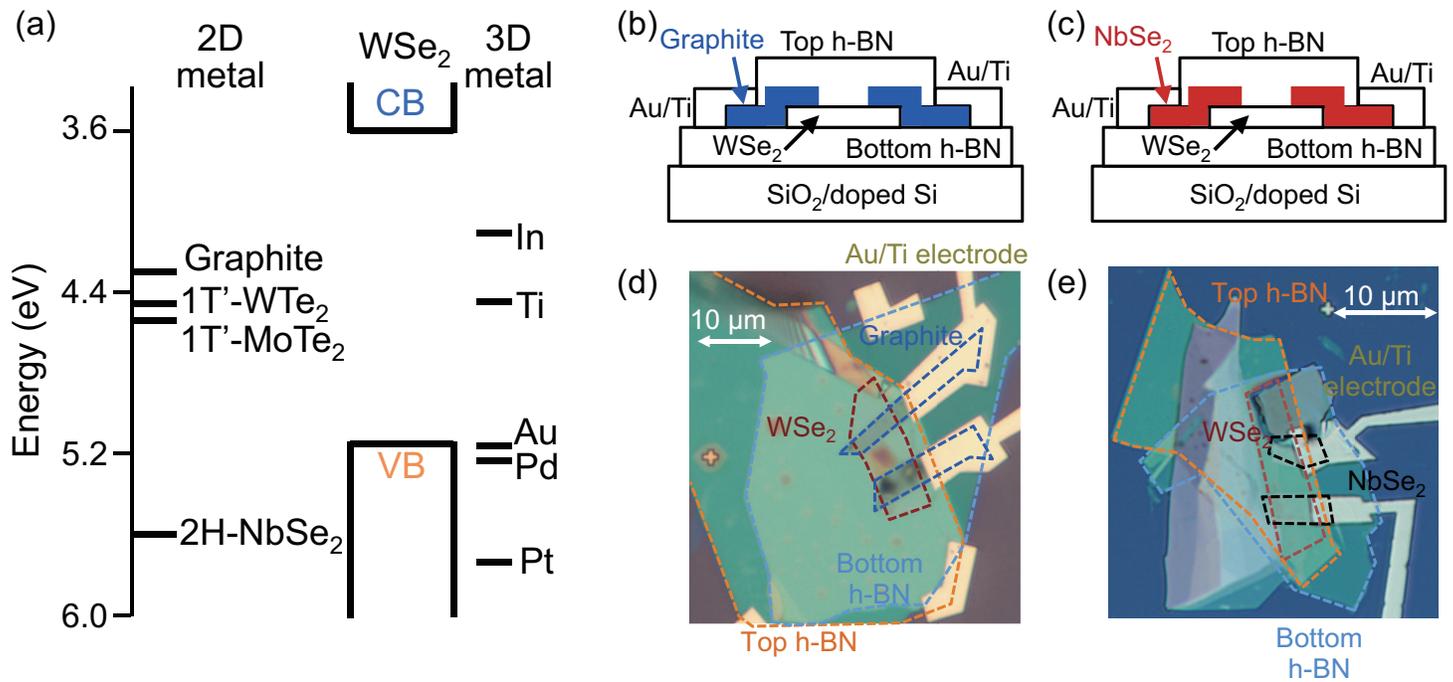

Figure 2

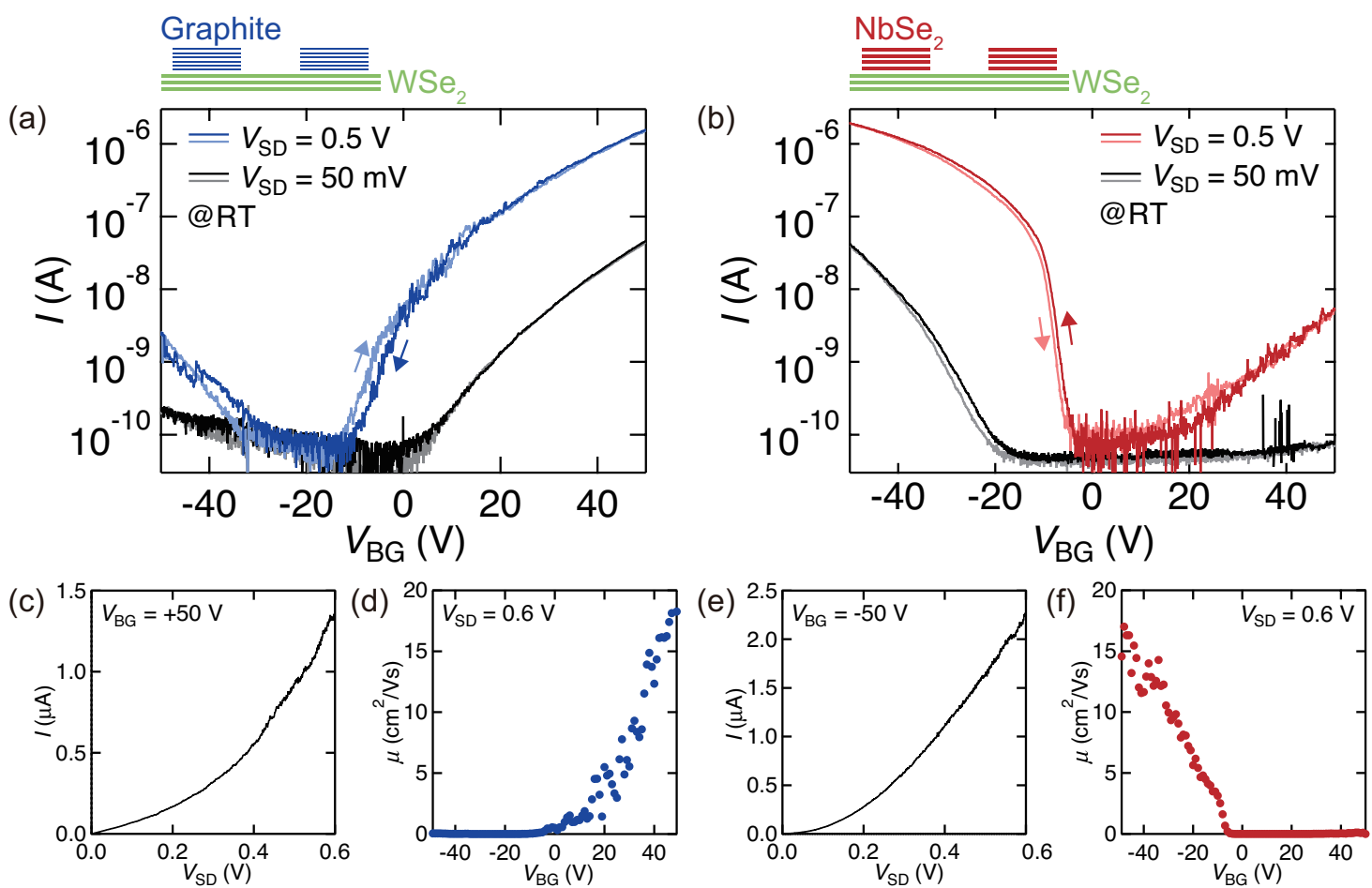

Figure 3

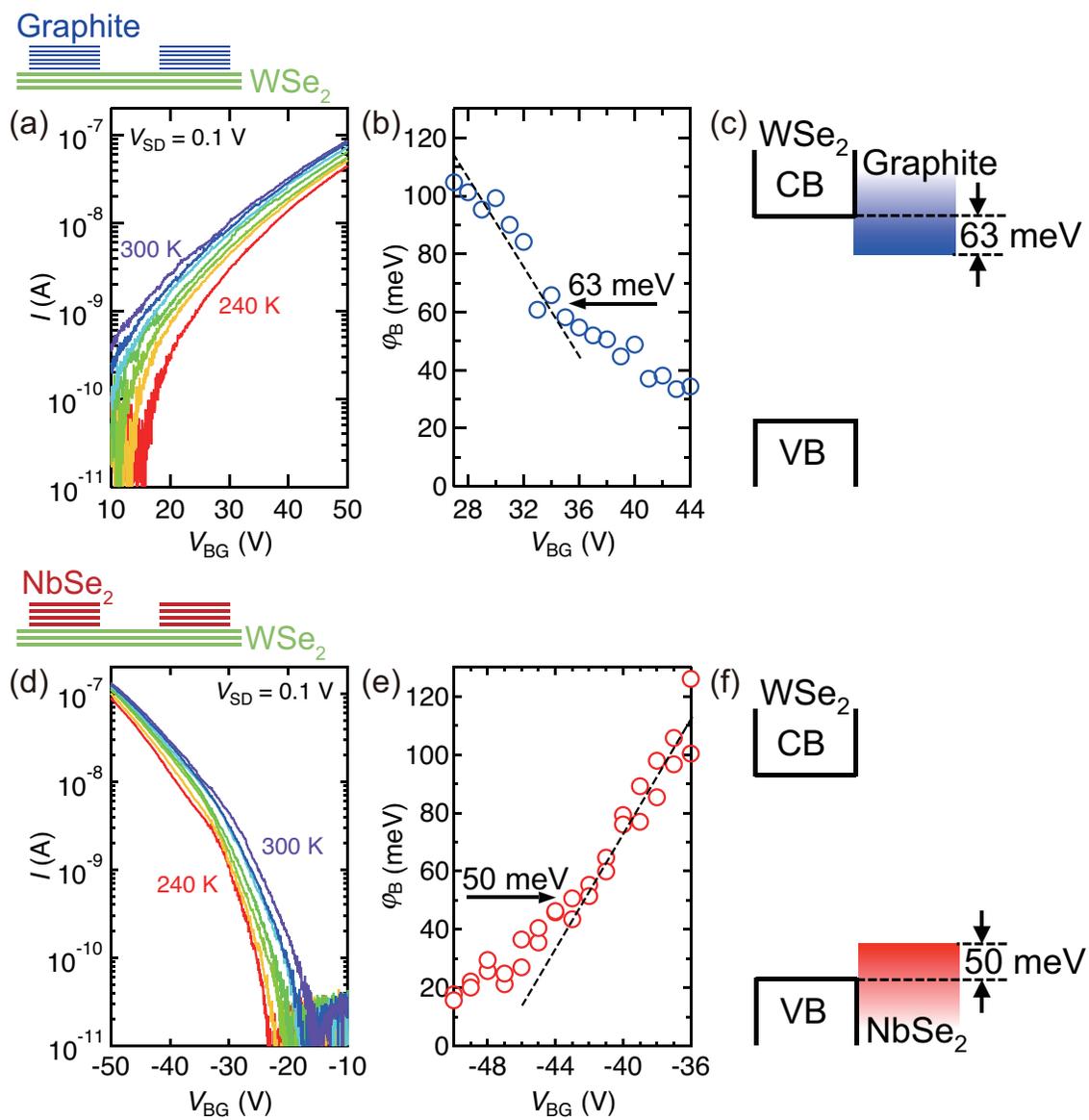